\documentclass[12pt,a4paper]{article}
\usepackage{amsthm}
\theoremstyle{plain}
\usepackage{authblk}
\usepackage[english]{babel}
\usepackage{graphicx}
\usepackage[latin1]{inputenc}
\usepackage{verbatim}
\usepackage{amsfonts}
\usepackage{amsmath}
\usepackage{txfonts}
\usepackage[T1]{fontenc}
\usepackage{color}
\usepackage{epsfig}
\usepackage{natbib}
\usepackage{url}
\usepackage[bookmarksnumbered=true, colorlinks=true, citecolor=blue]{hyperref}
\date{}

\title{General Covariance, Diffeomorphism Invariance, and Background Independence in 5 Dimensions}
\author{Antonio Vassallo}
\affil{University of Lausanne, Department of Philosophy, CH-1015 Lausanne\\ \url{antonio.vassallo@unil.ch}}

\begin{document}

\maketitle
\begin{center}
To appear in T. Bigaj and C. W\"uthrich (eds.): \emph{Metaphysics in Contemporary Physics}, pp. 229-250. Rodopi, 2015.\\
\end{center}
\pdfbookmark[1]{Abstract}{abstract}
\begin{abstract}
The paper considers the ``GR-desideratum'', that is, the way general relativity implements general covariance, diffeomorphism invariance, and background independence. Two cases are discussed where $5$-dimensional generalizations of general relativity run into interpretational troubles when the GR-desideratum is forced upon them. It is shown how the conceptual problems dissolve when such a desideratum is relaxed. In the end, it is suggested that a similar strategy might mitigate some major issues such as the problem of time or the embedding of quantum non-locality into relativistic spacetimes.\\

\end{abstract}

\textbf{Notation:} In the following, Einstein's convention will apply. Lower-case indices will be taken to range from $0$ to $3$, while upper-case ones will range from $0$ to $4$. A semicolon before one or more indices will indicate covariant differentiation with respect to those indices, a comma will instead indicate ordinary differentiation. Moreover, all equations will be written in natural units such that c=G=1.

\section{Introduction}\label{se1}
General relativity (GR) is the most corroborated theory of gravitation we have so far. It is also considered a spacetime theory because it is taken to unify gravitational phenomena and physical geometry.\footnote{I will not even try to address here questions regarding the meaning (if any) of the unification involved. The interested reader can take a look, for example, at \citet{50,394}.} The dynamics of GR is encoded in the Einstein's field equations:\footnote{I choose to disregard the cosmological constant, since it does not affect the analysis carried out in this paper.}
\begin{equation}\label{efe1}
\mathbf{G}[\mathbf{g}]=8\pi\mathbf{T}[\mathbf{g},\boldsymbol{\phi}].
\end{equation}
The left-hand side of the equation is the Einstein tensor defined over a $4$-di\-mensional differentiable semi-Riemannian manifold $M_{4}$: since it depends on the $4$-metric tensor $\mathbf{g}$, it is taken to encode information on the $4$-geometry of physical spacetime. The right-hand side comprises the stress-energy tensor, which encodes information such as the energy-momentum density of some matter field $\boldsymbol{\phi}$ distributed over (a region of) spacetime. A solution (or model) of (\ref{efe1}) is then a triple $<M_{4},\mathbf{g},\mathbf{T}>$ that represents a physical scenario where the geometry of spacetime (aka the gravitational field) interacts with matter fields distributed over it. In short: <<Space acts on matter, telling it how to move. In turn, matter reacts back on space, telling it how to curve.>> (\citealp[][p.5]{27}).\\
Equations (\ref{efe1}) are formulated in terms of geometrical field-theoretic objects, more precisely tensor fields, defined over a $4$-manifold $M_{4}$. In order to make them simpler to handle for sake of calculation, it is possible to introduce some coordinate system $\{x^{i}\}$ over a neighborhood $U$ of $M_{4}$, and then rewrite them in terms of components of the geometrical objects in that coordinate system:
\begin{equation}\label{efe2}
G_{ij}[g_{ij}]=8\pi T_{ij}.
\end{equation}
Unlike (\ref{efe1}), the equations compactly written as (\ref{efe2}) involve derivatives and symmetric matrices, which simplify (but not \emph{over}-simplify) the work of the physicists dealing with calculations. The important fact is that, by construction, the physical information encoded in the relations between field-theoretic objects as described by (\ref{efe1}) remain unchanged when switching to (\ref{efe2}), independently of the particular coordinatization chosen. For example, for two coordinate systems $\{x^{i}\}$ and $\{\tilde{x}^{i}\}$ and two associated bases $\{e^{i}\}$ and $\{\tilde{e}^{i}\}$ on a neighborhood of a point $P\in M_{4}$, we have:
\begin{equation}\label{chg}
\mathbf{g}=g_{ij}|_{P}e^{i}\otimes e^{j}=\tilde{g}_{ij}|_{P}\tilde{e}^{i}\otimes\tilde{e}^{j},
\end{equation}
the (invertible) transformation law relating $g_{ij}$ and $\tilde{g}_{ij}$, being:
\begin{equation}\label{glaw}
g_{ij}\rightarrow\tilde{g}_{ij}=\frac{\partial x^{h}}{\partial\tilde{x}^{i}}\frac{\partial x^{k}}{\partial\tilde{x}^{j}}g_{hk}.
\end{equation}
Since we know from elementary tensor calculus that (\ref{chg}) holds for whatever tensorial object, it follows that it is possible to recover (\ref{efe1}) from (\ref{efe2}) for whatever coordinate system $\{x^{i}\}$ defined on a neighborhood of a point in the manifold.\\
Further, it can be shown that to any (sufficiently smooth) coordinate transformation $\{x^{i}\}\rightarrow\{\tilde{x}^{i}\}$ defined in (a neighborhood of) $M_{4}$ corresponds a self-diffeomor\-phism\footnote{A self-diffeomorphism (from now on called simply diffeomorphism), is a mapping $f:M_{4}\rightarrow M_{4}$ that is bijective, continuous, and differentiable together with its inverse $f^{-1}$.} $f$ such that, for each point $P$ in (a neighborhood of) $M_{4}$, it is the case that $x^{i}(f(P))=\tilde{x}^{i}(P)$.\footnote{Suggestively speaking, while in the ``coordinate'' case we ``keep P fixed'' and we evaluate it (in the sense of associating a 4-tuple of real numbers to it) under a coordinate system \emph{different} from the starting one, in the ``diffeomorphic'' case we ``move P around the manifold'' and then we evaluate it under the \emph{same} coordinate system. For the reader unsatisfied by this, indeed, very rough characterization, \citet{12} has a section that nicely explains the difference between extrinsic (or passive) transformations expressed in terms of coordinate systems, and intrinsic (or active) transformations expressed in terms of diffeomorphisms.}\\
In the following, we will take advantage of the duality between the geometrical or intrinsic formulation of GR (involving geometrical objects and diffeomorphisms) and its local formulation (involving components of geometrical objects and change of coordinate systems), and we will switch back and forth from the intrinsic to the coordinates-related language for the sake of simplicity, according to the specific circumstances considered.\\
Nowadays, GR is so widely and firmly accepted by the scientific community to be considered one of the pillars of modern physics. At the root of such a vast agreement lies not only the empirical success of the theory, but also its numerous applications to, just to mention two major fields, astrophysics and cosmology. In the words of Misner, Thorne, and Wheeler:
\begin{quote}
Einstein's theory attracts the interest of many today because it is rich in applications. No longer is the attention confined to three famous but meager tests: the gravitational redshift, the bending of light by the sun, and the precession of the perihelion of Mercury around the sun. The combination of radar ranging and general relativity is, step by step, transforming the solar-system celestial mechanics of an older generation to a new subject, with a new level of precision, new kinds of effects, and a new outlook. Pulsars, discovered in 1968, find no acceptable explanation except as the neutron stars predicted in 1934, objects with a central density so high $(\sim 10^{14} g/cm^{3})$ that the Einstein predictions of mass differ from the Newtonian predictions by $10$ to $100$ per cent. About further density increase and a final continued gravitational collapse, Newtonian theory is silent. In contrast, [GR] predicted [...] the properties of a completely collapsed object, a ``frozen star'' or ``black hole''.\\
\citep[][pp. viii,ix]{27}
\end{quote}
However, the success of GR is not only due to its empirical adequacy and its richness in applications. Many physicists (and mathematicians) would in fact add that GR is an extremely elegant and mathematically beautiful theory. This further virtue of the theory should not be regarded as merely aesthetic, but it should be taken as a hint of the fact that GR points at some deep physical ``truth'' about the world. Dirac addresses this view as follows:
\begin{quote}
Let us now face the question, suppose a discrepancy had appeared, well confirmed and substantiated, between the theory and observations. [...] Should one then consider the theory to be basically wrong? [...] I would say that the answer to the [q]uestion is emphatically NO. The Einstein theory of gravitation has a character of excellence of its own. Anyone who appreciates the fundamental harmony connecting the way Nature runs and general mathematical principles must feel that a theory with the beauty and elegance of Einstein's theory \emph{has} to be substantially correct. If a discrepancy should appear in some applications of the theory, it must be caused by some secondary feature relating to this application which has not been adequately taken into account, and not by a failure of the general principles of the theory.
\citep[][pp. 6,7 (Dirac's emphasis)]{400}
\end{quote}
The questions about how, in general, a physical theory becomes accepted by the community, and what features should be privileged over others when assessing the merits of such a theory (e.g. empirical over formal), are beyond the scopes of this essay (see, for example, \citealp{399} for a historical reconstruction and a philosophical discussion of the reasons that led to the acceptance of Einstein's theory). Here we will just make the working hypothesis that there is ``something'' physical \emph{and} formal about GR (what Dirac calls the <<general principles of the theory>>) that, according to mainstream physicists, places a constraint on future generalizations of the theory (to include, for example, quantum effects). In the remainder of the section, we will briefly consider some candidates for this ``something'', and we will finally propose a tentative definition that captures the <<character of excellence>> of GR that should therefore be preserved in any decent generalization of the theory.\\
To start our inquiry, let us consider the most straightforward candidate to be such ``something'': general covariance. With the designation ``general covariance'' here we will intend the formal invariance of the dynamical equations (\ref{efe2}) under an arbitrary coordinate transformation. Einstein originally thought that general covariance so intended was some sort of generalization of the principle of relativity that holds in Newtonian mechanics and special relativity. Very simply speaking, while in these two latter theories the description of the dynamics of a physical system remained unchanged under coordinate transformations between inertial reference frames, so that these frames where considered physically equivalent,\footnote{The question of the equivalence between coordinate systems and frames of reference is another extremely delicate matter I will not touch upon here. See, e.g., \citet[][section 6.3]{365}.} in GR such a description is unchanged under whatever coordinate transformation, so one might claim that \emph{all} reference frames are physically equivalent. Can the ``virtue'' of GR reside in its general covariance simpliciter?\\
The answer, as firstly pointed out  by \citet{59}, is no since it is possible to render generally covariant also a theory that patently accords a privileged status to inertial reference systems. Hence, general covariance cannot be taken as the implementation of the physical requirement that all reference frames are physically equivalent. To understand why it is so, consider a theory of a massless Klein-Gordon field $\phi$ over a manifold endowed with a Minkowski metric $\boldsymbol{\eta}$. We can write the field equations of the theory as:
\begin{equation}\label{kg0}
\eta^{ij}\phi_{,ij}=0.
\end{equation}
Equation (\ref{kg0}) is not generally covariant because it holds only in inertial coordinate systems. 
In order to make it generally covariant we just have to make the purely formal move of rewriting it in a way such that it preserves its form in whatever coordinate system \emph{and} reduces to (\ref{kg0}) in those coordinate systems that are inertial.\\
This means that we have to write $\boldsymbol{\eta}$ in a generalized form $\mathbf{g}$ and introduce the (unique) compatible covariant derivative operator $\boldsymbol{\nabla}$,\footnote{``Compatible'' means that $\boldsymbol{\nabla}\mathbf{g}=0$.} so that (\ref{kg0}) finally reads:
\begin{equation}\label{kg1}
g^{ij}\phi_{;ij}=0.
\end{equation}
It is easy to understand that (\ref{kg1}) is invariant in form under whatever coordinate transformation but, still, the theory - being specially relativistic - accords to inertial frames of reference a privileged status. In the language of geometrical objects,  (\ref{kg1}) is written
\begin{equation}\label{kg2}
\Box_{\mathbf{g}}\phi=0,
\end{equation}
and, as expected, it is invariant in form (that is, covariant) under diffeomorphic transformations. Since such a move can be performed for whatever spacetime theory - even Newtonian mechanics - then (i) general covariance per se has no physical import and (ii) the generally covariant nature of GR cannot be linked to a peculiar feature such as a generalized principle of relativity. If Kretschmann's point is correct, then what is it that renders GR such a peculiar and successful theory?\\
The most simple reaction would be to claim that the peculiarity of GR does not reside just in the mere formal invariance of its field equations, but also in the physical import of the solutions space of such equations. If, in fact, we consider two solutions of (\ref{kg1})/(\ref{kg2}) related by a diffeomorphism $f$, they will represent one and the same physical situation if and only if $f$ is an isometry\footnote{Roughly speaking, $f\in diff(M_{4})$ is an isometry for a geometrical object $\boldsymbol{\theta}$ if the application of such a transformation to the object does not change it: $f^{*}\boldsymbol{\theta}=\boldsymbol{\theta}$.} of the Minkowski metric. Not surprisingly, the group of isometries $iso(\boldsymbol{\eta})$ of the Minkowski metric is just the Poincar\'e group. In the case of equation (\ref{efe1}), instead, since the metric $\boldsymbol{g}$ is not fixed a priori but it is subject to the dynamical evolution, the group $diff(M_{4})$ is unrestricted and, as expected, whatever two solutions of (\ref{efe1}) related by a diffeomorphism are physically indistinguishable.\footnote{Just to have a more concrete idea of what this means, consider a solution $\mathfrak{M}$ of (\ref{efe1}) where we can construct a physical observable interpreted as a $4$-distance function $d_{\mathbf{g}}:M_{4}\times M_{4}\to\mathbb{R}$ defined in terms of the metric $\mathbf{g}$. Let us now take a diffeomorphically transformed model $f^{*}\mathfrak{M}$, where $f\in diff(M_{4})$; in such a model we can construct an observable diffeomorphically related to $d_{\mathbf{g}}$, namely a function $d_{f^{*}\mathbf{g}}$ defined by  the diffeomorphic metric $f^{*}\mathbf{g}$. It is now easy to see that, taken two arbitrary points $P,Q\in M_{4}$, we have $d_{\mathbf{g}}(P,Q)=d_{f^{*}\mathbf{g}}(f(P),f(Q))$. It follows that no $4$-distance measurement can discriminate $\mathfrak{M}$ from $f^{*}\mathfrak{M}$. By repeating the same reasoning for all the observables constructible in $\mathfrak{M}$ and $f^{*}\mathfrak{M}$, we can conclude either that the two models represent different physical possibilities which are empirically indistinguishable - and then we would immediately fall pray of arguments from indeterminism, such as the infamous ``hole argument'', cf. \citet{11} - or that the two models represent one and the same physical scenario.\label{det}}\\
The above reply points at the fact that what makes the fortune of GR is not just its general covariance, but its diffeomorphism invariance: in virtue of having a dynamical geometry, GR has no fixed isometries different from the whole $diff(M_{4})$ and, hence, diffeomorphisms are physical symmetries of the theory. Indeed, $diff(M_{4})$ can be considered the \emph{gauge group} of GR.\\
The question whether GR can be really considered a gauge theory in the sense commonly intended in particle physics is complex and highly debated (to have a substantive example of such debate, see \citealp{174,370}); here we will just sketch a counter-reply to the above argument.\\
If the real difference between GR and previous spacetime theories is the dynamical status of spacetime's geometry, then we can just straightforwardly rewrite whatever spacetime theory by adding dynamical constraints on the spatiotemporal structures, so that such a difference can be obliterated. To see this, let us take the above theory of a Klein-Gordon field over a Minkowski spacetime and rewrite the dynamical equations as:
\begin{subequations}\label{kg3}
\begin{equation}\label{suk1}
\mathbf{Riem}[\mathbf{g}]=0,
\end{equation}
\begin{equation}\label{suk2}
\Box_{\boldsymbol{g}}\phi=0,
\end{equation}
\end{subequations}
$\mathbf{Riem}[\mathbf{g}]$ being the Riemann curvature tensor. The Minkowski metric, now, is a dynamical object picked up as a solution of (\ref{suk1}). The theory whose dynamics is encoded in (\ref{kg3}) has the feature that two models related by whatever diffeomorphism  represent in fact one and the same physical situation.\\
The immediate reaction to such an example is that it exploits a mere mathematical trick to render the transformations in $diff(M_{4})\setminus iso(\boldsymbol{\eta})$ just \emph{mathematical} symmetries of the theory and then mix them up to the \emph{physical} symmetries in $iso(\boldsymbol{\eta})$ in order to patch them together into a fake gauge group. 
In this particular case, a possible implementation of this counter-argument is, following \citet{66,15}, to point out that some geometrical objects have been artificially rendered dynamical and, thus, the theory can be deparametrized by quotienting out the redundant structures in the solutions space introduced by adding the fake dynamical sector.\footnote{Just to have a rough idea of what adding redundant structure means when passing from (\ref{kg2}) to (\ref{kg3}), it is sufficient to point out that to each solution $<M_{4},\mathbf{g},\phi>$ of (\ref{kg2}) corresponds a set of solutions $\{<M_{4},\mathbf{g},\phi>,<M_{4},f^{*}\mathbf{g},f^{*}\phi>,<M_{4},h^{*}\mathbf{g},h^{*}\phi>,\dots\}$ of (\ref{kg3}) with $f,h,\dots\in diff(M_{4})\setminus iso(\boldsymbol{\eta})$.} Such a strategy makes sense in the present case: if we deparametrize (\ref{kg3}) by quotienting out the solution space by $diff(M_{4})\setminus iso(\boldsymbol{\eta})$, we end up again with (\ref{kg2}). If this reasoning could be carried out for all spacetime theories prior to GR, we would then have a reasonable candidate to be the peculiar feature of this latter theory, namely, the absence of spacetime structures that are not genuinely dynamical. Some authors would dub such a feature ``background independence''.\\
However, the above reasoning cannot be the final word in the debate. The present case per se is, in fact, too artificial to represent a really difficult challenge. Here the Minkowski metric represents an elephant in the room in the sense that we are perfectly able to see that such a structure ``persists'' unaltered in all models of the theory modulo a diffeomorphism transformation.\footnote{Which means that, for whatever two models $<M_{4},\mathbf{g},\phi>$ and $<M_{4},\mathbf{g'},\phi'>$, there is \emph{always} some $f\in diff(M_{4})$ such that $\mathbf{g'}=f^{*}\mathbf{g}$.} In general, however, making the distinction between genuine dynamical and pseudo-dynamical structures varies from difficult to impossible so that defining background independence in an accurate manner is a hard task (\citealp{354,47}, discuss at length the problem; see also \citealp{311}  for an alternative account of background independence).\\
So, in the end, what is it that confers on GR a <<character of excellence of its own>>? Let us try to filter from the above discussion the salient uncontroversial claims regarding the features of GR, and come up with the following tentative characterization: Unless its predecessors, GR is a $diff(M_{4})$-gauge theory whose only non-dynamical constraint is to be formulated over some $4$-dimensional semi-Riemannian manifold, and that does not postulate more spatiotemporal structure than just that encoded in a Lorentzian $4$-metric. If this is (or gets close to) what is at the root of the major success of GR, then we can consider it as a (minimal) constraint on any future (quantum) spacetime theory that seek to supersede GR:

\newtheorem*{fed}{GR-desideratum}
\begin{fed}
Any theory which is a valid candidate for superseding GR must be [become in the classical limit] a $diff(M_{d})$-gauge theory whose only non-dynamical constraint is to be formulated over some $d$-dimensional semi-Riemannian manifold and that does not postulate more spatiotemporal structure than just that encoded in a $d$-metric.\footnote{The $g_{ij}$-part of such a metric having Lorentzian signature.}
\end{fed}

Note that, in general, besides arguments from physics, the GR-desideratum can be backed up by metaphysical arguments. One might for example argue that (\ref{efe1}) is preferable to (\ref{kg3}) because, in this latter theory, there is a ``lack of reciprocity'' between the Minkowski metric and the scalar field: in (\ref{kg3}) spacetime tells matter how to move, but matter does not tell spacetime how to curve.\\
In the following, I will show how a too strict pursuit of the GR-desideratum might negatively affect the interpretation of prima facie genuine generalizations of GR.

\section{5-Dimensional Extensions of General Relativity}\label{se2}
The Kaluza-Klein approach, so called because, historically, the first attempts in this sense were made by Theodor \citet{30} and, slightly later, by Oskar \citet{31}, started as a theoretical program that sought to unify gravitational and electromagnetic forces as curvature effects of a $5$-dimensional semi-Riemannian manifold. This is achieved by considering the $5$-dimensional vacuum Einstein's equations:\footnote{In the $4$-dimensional case, if we take (\ref{efe1}) and we set the stress energy tensor equal to zero (that is, no matter is present), we obtain that the equations reduce to $\mathbf{Ric}[\mathbf{g}]=0$, $\mathbf{Ric}$ being the so-called Ricci tensor. This is why they are in general said to be ``vacuum'' field equations.}
\begin{equation}\label{5fe}
R_{AB}[g_{AB}]=0,
\end{equation}
with the $5$-Ricci tensor $R_{AB}$ depending on a metric of the form:
\begin{equation}\label{KKm}
g_{AB}=
\left( \begin{array}{cc}
g_{ij}+\kappa^{2}\phi^{2}A_{i}A_{j} & \kappa^{2}\phi^{2}A_{i}\\
\kappa^{2}\phi^{2}A_{j} & \phi^{2}\\
\end{array} \right),
\end{equation}
where $\mathbf{A}$ is the $4$-vector potential, $\phi$ is a scalar field, and $\kappa$ is a free parameter. The signature of the $5$-metric will be assumed to be Lorentzian, thus identifying the extra fifth dimension as space-like.\\
The original Kaluza-Klein theory placed two very strong constraints on the fifth dimension, namely, (i) that all partial derivatives with respect to the fifth coordinate are zero (cylinder condition), and (ii) that the fifth dimension has a closed short-scale topology (compactification condition).\footnote{To be fair to historians of physics, (i) was proposed by Kaluza, while (ii) was added later by Klein, who further showed that, in fact, (ii) implies (i).} The most important consequences of these conditions are that there is no change in $4$-dimensional physical quantities that can be ascribed to the presence of an extra spatial dimension and that such a fifth dimension is unobservable at low energies. Condition (ii) was also a vital ingredient in the attempt to explain the quantization of electric charge.\footnote{Attempt which experiments proved unsuccessful: According to Kaluza-Klein theory, the electron had a mass of twenty-two orders of magnitude higher then the measured one.} En passant, it is interesting to note that condition (ii) obviously prevents whatever macroscopic object from spanning the fifth dimension.\footnote{The reader is invited to think about the strange consequences for spatial $3$-dimensional beings like us of living in an uncompactified $4$-space. For example, if we were allowed to move in an extra spatial dimension, we could ``switch'' the left and right sides of our body without altering the up-down and front-back orientations. Moreover, from the perspective of the extra dimension, the interior of our bodies would be exposed (our skin represents just a $3$-dimensional boundary).}\\
If we set $\phi=1$ and $\kappa=4\sqrt{\pi}$ in (\ref{KKm}), and we substitute it in (\ref{5fe}), the $ij$-components of the field equations become:
\begin{equation}\label{grem}
G_{ij}=8\pi T_{ij}^{EM},
\end{equation}
with $G_{ij}$ the Einstein $4$-tensor and $T_{ij}^{EM}=\frac{1}{4}g_{ij}F_{kl}F^{kl}-F_{i}^{\phantom{i}k}F_{jk}$ the electromagnetic stress-energy $4$-tensor of standard GR. The i4-components, instead, become:
\begin{equation}\label{emgr}
F_{ij}^{\phantom{ij};i}=0,
\end{equation}
with $F_{ij}=A_{j,i}-A_{i,j}$ the Faraday $4$-tensor. In short, (\ref{5fe}) collapse into Einstein's field equations of GR and the (source free) Maxwell's equations of electromagnetism coupled together. This is the so called ``Kaluza-Klein miracle'' and represents a remarkable result in that it  ``geometrizes away'' the electromagnetic field as GR does with gravity, thus suggesting that both gravity and electromagnetism are just a manifestation of the geometry of a $5$-dimensional spacetime. Is this a right suggestion? Let us consider a reason why it does not seem so.\\
In order to see what is fishy about the interpretation of Kaluza-Klein theory as a genuine $5$-dimensional theory, we just have to notice that conditions (i) and (ii) restrict the possible topologies of the $5$-manifold to $M_{4}\times S^{1}$, with $M_{4}$ whatever topology compatible with the $ij$-part of the metric tensor (\ref{KKm}) and $S^{1}$ the topology of the circle. This hints at the fact that, as long as we restrict our attention to the $ij$-part of the theory, we are dealing with standard GR and, hence, whatever coordinate transformation $x^{i}\rightarrow \tilde{x}^{i}=f(x^{i})$ we apply to the metric (\ref{KKm}), we always end up with the same equations (\ref{grem}) \emph{and} we do not change the physical information originally encoded in the $4$-metric before the transformation. What about the components of the metric depending on the fifth coordinate? In this case, the freedom in choosing the coordinate transformations that leave (\ref{emgr}) invariant in form and do not alter the physical information encoded in (\ref{KKm}) is severely limited. In fact, the only possible choice that we can make is $x^{4}\rightarrow\tilde{x}^{4}=x^{4}+f(x^{i})$. To realize why it is so, we have just to plug this transformation in the $5$-dimensional extension of the law (\ref{glaw}) for the transformation of the metric tensor: it is easy to see that such a transformation induces just one change in the metric (\ref{KKm}), namely, $A_{i}\rightarrow\tilde{A}_{i}=A_{i}+\frac{\partial f(x^{i})}{\partial x^{i}}$, which obviously does not alter the electromagnetic part of the theory. In short, the gauge group of the theory is not $diff(M_{5})$ but $diff(M_{4})\times U(1)$.\\
If, now, we claim that a genuine $5$-dimensional extension of GR should admit $diff(M_{5})$ as gauge group, we cannot but reach the conclusion that Kaluza-Klein theory is a fake $5$-dimensional theory in that it is a version of $4$-dimensional GR that camouflages the gauge group of electromagnetism by ``spatializing'' the symmetries of the theory using an additional space-like dimension: there is no substantial unification here, just a mere algebraic play.\footnote{This line of reasoning can be found, for example, in \citet[][p.87]{16}.}
Note that the above reasoning is based on the simple fact that the conditions of cilindricity and compactification are fixed a priori, i.e., they are not part of the dynamics of the theory. Simply speaking, the theory is not really dealing with solutions of (\ref{5fe}) but with a small subset of it, namely Ricci-flat spacetimes with topology $M_{4}\times S^{1}$ where, additionally, 
$S^{1}$ has to ``appear'' at short scales. This situation is nothing but a more complex instance of that occurring with the theory (\ref{kg1})/(\ref{kg2}). It is then clear that we can in principle modify the field equations (\ref{5fe}) so that they can encode cilindricity and compactification conditions\footnote{For example, by adding some conditions that pick up at each point of $4$-spacetime a fifth-dimensional-like vector field whose integral curves are circles with small radii.} in order to have a fully $diff(M_{5})$-invariant theory, exactly as (\ref{kg3}) is a $diff(M_{4})$-invariant extension of (\ref{kg1})/(\ref{kg2}).\\
Such a move is exactly as controversial as the one discussed in the previous section, but can it be charged of hiding an elephant in the room as it was the case with the Minkowski metric in (\ref{kg3})? Well, if we hold the firm commitment to the GR-desideratum, then we can clearly hear the $M_{4}\times S^{1}$ structure trumpeting beneath the dynamics. Notice, however, that in this case the GR-desideratum cannot be backed up with compelling metaphysical arguments as, for example, the ``reciprocity'' one mentioned at the end of the previous section. Here we are dealing with an empty spacetime, so there is nothing acting on anything without being affected in return. Hence it would not be that weird to claim that the dynamically extended version of Kaluza-Klein theory (\ref{5fe}) just depicts spacetimes that are, \emph{as a bare matter of fact}, more structured then a generic (empty) general relativistic spacetime, in that it adds to such generic spacetime an extra-dimensional tiny ring for each point. The counter-reply would be that, in this case, such a theory would be unable to explain \emph{why} spacetime has to have such a structure. This is fair enough, but any theory has to start from some assumptions in order to be developed: even standard GR does not explain why spacetime has to be $4$-dimensional and Lorentzian; it just takes as a bare matter of fact that special relativity holds at quasi-point-sized regions. The whole discussion, then, seems to turn into metaphysical tastes about what has to be counted as a bare fact and what has not. We could, for example, endorse some variant of the regularity account of laws where the dynamical laws of Kaluza-Klein theory just supervene on a pure spatiotemporal Humean mosaic of local matters of particular fact. In such a context, asking why the Kaluza-Klein $5$-spacetime has this structure is a misleading question: it just happen to show a pattern of regularities that is best described by the equations of the theory.\\
Let us recap: if we insists on the GR-desideratum, then we are driven to judge Kaluza-Klein theory as a mathematical trompe-l'\oe il that spatializes the symmetries of the electromagnetic theory, so it cannot count as a genuine unification of gravity and electromagnetism. However, such a judgment seems too harsh and does not involve in any way the empirical adequacy of the theory (whose failure is the only uncontroversial argument that historically led to the dismissal of Kaluza-Klein theory). It is in fact possible to make perfect physical and metaphysical sense of Kaluza-Klein theory as a unified theory by relaxing the GR-desideratum.\\
The moral to draw from the discussion so far is that, all things considered, perhaps the GR-desideratum is not that strict a desideratum: we can relax such a requirement without necessarily falling into patently artificial theories such as (\ref{kg1})/(\ref{kg2}). But let us go ahead in the analysis, and show a case where the GR-desideratum is not only too rigid, but has also metaphysically odd consequences.\\
Since the strongest source of resistance to the idea that Kaluza-Klein theory represents a genuine $5$-dimensional theory lies in the conditions of cilindricity and compactification, let us now consider an implementation of the Kaluza-Klein program\footnote{See \citealp{43} for an extensive survey of Kaluza-Klein gravity.} that dispenses with such constraints. This is the case of induced matter theory, especially known in its ``space-time-matter'' version as firstly put forward by \citet{34}. This theory (or, better, class of theories) rests on the mathematical result that any analytic $N$-dimensional [semi-] Riemannian manifold can be locally embedded in a $(N+1)$-dimensional Ricci-flat [semi-] Riemannian manifold (Campbell-Magaard theorem, see for example \citealp{33}); hence, the field equations of the theory are, again, (\ref{5fe}).\\
If we now write a $5$-metric in its diagonal form:
\begin{equation}\label{indmet}
g_{AB}=
\left( \begin{array}{ccccc}
\phantom{1} & \phantom{1} & \phantom{1} & \phantom{1} & 0\\
\phantom{1} & \phantom{1} & \phantom{1} & \phantom{1} & 0\\
\phantom{1} & g_{ij} & \phantom{1} & \phantom{1} & 0\\
\phantom{} & \phantom{1} & \phantom{1} & \phantom{1} & 0\\
0 & 0 & 0 & 0 & g_{44}\\
\end{array} \right),
\end{equation}
and we substitute it in (\ref{5fe}), we obtain again the $4$-dimensional Einstein's field equations plus an expression for the $4$-stress-energy tensor of the form (see \citealp{35}, for the detailed calculations):
\begin{equation}\label{strs}
T_{ij}=\frac{(\phi_{,i})_{;j}}{\phi}-\frac{1}{2\phi^{2}}\Bigg\{\frac{\phi_{,4}g_{ij,4}}{\phi}-g_{ij,44}+g^{kl}g_{ik,4}g_{jl,4}-\frac{g^{kl}g_{kl,4}g_{ij,4}}{2}+\frac{g_{ij}}{4}\Big[g^{kl}_{\phantom{kl},4}g_{kl,4}+(g^{kl}g_{kl,4})^{2}\Big]\Bigg\},
\end{equation}
where, for notational simplicity, it is assumed $g_{44}=\phi^{2}$.\\ 
From these constructions it follows that, whenever we consider a $4$-hypersurface $\Sigma_{4}$ by fixing $x^{4}=const.$, we obtain a $4$-metric $g_{ij}$ and a stress-energy tensor $T_{ij}$ both well-defined on $\Sigma_{4}$.\\
Just to have a rough idea of how this works, let us consider an example taken from \citet{32} and, in order to simplify the notation, let us follow the author in naming the coordinates as follows: $x^{0}=t, x^{1}=x, x^{2}=y, x^{3}=z, x^{4}=\psi$. In case of a homogeneous and isotropic $5$-dimensional universe, a class of $5$-line elements (parametrized by $\alpha\in\mathbb{R}_{\ne 0}$) correspondent to $5$-metrics which are solutions to (\ref{5fe}) is:
\begin{equation}\label{frw5}
d\mathfrak{s}^{2}=\psi^{2}dt^{2}-t^{\frac{2}{\alpha}}\psi^{\frac{2}{1-\alpha}}(dx^{2}+dy^{2}+dz^{2})-\frac{\alpha^{2}}{(1-\alpha)^{2}}t^{2}d\psi^{2}.
\end{equation}
It is easy to see that, if we restrict to $4$-hypersurfaces ($\psi=const., d\psi=0$), (\ref{frw5}) is reduced to:
\begin{equation}\label{frw4}
ds^{2}=dt^{2}-R^{2}_{\alpha}(t)(dx^{2}+dy^{2}+dz^{2}),
\end{equation}
which is just the family of Friedman-Lemaître-Robertson-Walker metrics in $4$-dimensions corresponding to flat $3$-geometries. Moreover, if we substitute (\ref{frw5}) in (\ref{strs}), we recognize that the stress-energy tensor resembles that of a perfect fluid with density $\rho$ and pressure $p$ given by:
\begin{subequations}\label{eqst}
\begin{equation}
\rho=\frac{3}{8\pi(\alpha\psi t)^{2}},
\end{equation}
\begin{equation}
p=\Bigg(\frac{2\alpha}{3}-1\Bigg)\rho.
\end{equation}
\end{subequations}
Hence, by fixing $\alpha$ and $\psi$ we obtain different possible FLRW $4$-models, with state equation for matter given by (\ref{eqst}). For example, for $\alpha=\frac{3}{2}$, we would have a $4$-dimensional universe filled with dust, while for $\alpha=2$ we would have one filled with radiation.\\
The moral to be drawn from the above facts is that matter and energy are $4$-dimensional physical properties of spacetime ``induced'' by the $5$-metric tensor. Therefore, the cosmological solutions of (\ref{5fe}) all agree in giving us a picture of a $5$-dimensional empty universe whose restriction to a $4$-surface is a spacetime curved by matter.\\
In conclusion, it seems that now we have a truly $5$-dimensional generalization of GR: the theory is $diff(M_{5})$-invariant and there are neither background geometrical objects camouflaged as dynamical, nor bizarre restrictions on the topology of $5$-spacetime. However, such a complete fulfillment of the GR-desideratum does not come for free.\\
To see what is the price to be paid, consider, for example \citep{186}, a $5$-dimensional Minkowski spacetime. The correspondent line element in polar coordinates reads:
\begin{equation}\label{5min}
d\mathfrak{s}^{2}=dt^{2}-dr^{2}-r^{2}d\Omega^{2}-d\psi^{2},
\end{equation}
whose sections $\psi=const.$ are $4$-dimensional Minkowski spacetimes.\\
However, if we consider the transformation:
\begin{equation}\label{tru}
t'=t,\quad r'=\frac{r}{\psi}\Big(1+\frac{r^{2}}{\psi^{2}}\Big)^{-\frac{1}{2}},\quad\psi'=\psi\Big(1+\frac{r^{2}}{\psi^{2}}\Big)^{\frac{1}{2}},
\end{equation}
then (\ref{5min}) becomes
\begin{equation}\label{5frw}
d\mathfrak{s}^{2}=dt'^{2}-\psi'^{2}\Big(\frac{dr'^{2}}{1-r'^{2}}+r'^{2}d\Omega^{2}\Big)-d\psi'^{2},
\end{equation}
whose sections $\psi'=const.$ are FLRW spacetimes.\\
Using the intrinsic language, we can think that (\ref{5min}) belongs to a $5$-model $\mathfrak{M}=<M_{5},g_{AB}>$, and that the transformation (\ref{tru}) is dual to a $5$-diffeomorphism $f$, such that the metric (\ref{5frw}) belongs to the diffeomorphic model $f^{*}\mathfrak{M}=<M_{5},f^{*}g_{AB}>$. We therefore have two $5$-models which are diffeomorphically equivalent, but that induce $4$-models, which are empty in the former case and non-empty in the latter. What is the physical meaning of this fact? None, and that is exactly the problem.\\
If, in fact, we claim that $f$ is a gauge transformation, then we cannot but accept that $\mathfrak{M}$ and $f^{*}\mathfrak{M}$ are physically indistinguishable, which means that (\ref{5min}) and (\ref{5frw}) carry exactly the same physical information, namely, that related to the 5-geometry of spacetime. The fact that we are free to convey such an information by writing the line element either in the form (\ref{5min}) or (\ref{5frw}) means that such freedom has no physical consequences: this is exactly what physicists refer to as ``gauge freedom''. The puzzling consequence is that it makes no physical difference whether we take the 5-spacetime under consideration as a ``pile'' of empty Minkowski 4-spacetimes or matter-filled FLRW 4-spacetimes, since, in the end, the only thing that counts for the theory is the 5-geometry of spacetime. Crudely speaking, the theory is ``blind'' to whatever change in 4-spacetimes (as long as their piling up leads to the same 5-spacetime), and if we insist that $\mathfrak{M}$ and $f^{*}\mathfrak{M}$ are ontologically distinct albeit physically indistinguishable, we commit the theory to indeterminism, as already mentioned in footnote \ref{det}. For let us assume for the sake of argument that the dynamics of the theory under scrutiny can be cast in a (4+1) fashion, that is, that a 5-geometry can be recovered by considering a generic foliation of 4-hypersurfaces, specifying some initial data on one of such surfaces, and then evolving these data on the subsequent leaves of the foliation. In this case, the gauge freedom of the theory - as hinted above - would be translated in the freedom of choosing such a foliation without altering the dynamical evolution. Now let us consider a foliation $\mathcal{F}$ whose folios are Minkowski 4-spacetimes and another one $\mathcal{F'}$ which is identical to $\mathcal{F}$ except for the fact that some leaves in-between are FLRW 4-spacetimes. If we claim that $\mathcal{F}$ and $\mathcal{F'}$ are ontologically distinct, then so will be the corresponding dynamical evolutions and, hence, we are forced to accept that, even by specifying with arbitrary precision the data on an initial Minkowski 4-surface (which is common to $\mathcal{F}$ and $\mathcal{F'}$), the theory is unable to single out a unique dynamical evolution between $\mathcal{F}$ and $\mathcal{F'}$, which means that the theory is indeterministic. Such a situation is of course avoided if we take $\mathcal{F}$ and $\mathcal{F'}$ to represent one and the same physical situation.\\
In conclusion, since the theory is $diff(M_{5})$-invariant, all the quantities that are not invariant under $5$-diffeomorphisms are just ``gauge fluff'' that can be fixed at will without changing the physical information conveyed by the theory. This is the case for all the $4$-dimensional quantities extracted by $5$-dimensional ones by fixing the gauge (for example by setting $\psi=const.$): under a different gauge fixing, in fact, we would obtain different $4$-quantities from the same $5$-ones. Of course, the $4$-dimensional stress-energy tensor (\ref{strs}) is among the non-gauge invariant quantities.\\
If we take seriously this picture, then we cannot but claim that, in this theory, matter is just an unphysical illusion which merely depends on the $4$-dimensional ``perspective'' from which $5$-dimensional spacetime is seen, just like a holographic sticker that shows a different image depending on how it is inclined with respect to the light source. One could say that the unreality of matter in this theory already stems from the fact that, at the fundamental level, there is nothing but $5$-spacetime, but that would be a half-truth. While in fact we can agree that according to equations (\ref{5fe}) \emph{at the fundamental ontological level} there is nothing but $5$-dimensional geometry, this does not rule out the possibility that matter configurations \emph{supervene} on such a geometry, and hence are real albeit ontologically non-fundamental.\\
Naively, we could solve such a ``problem of matter'' just by declaring our firm pre-theoretical commitment to the existence of matter as a substance and hence arguing for the dismissal of induced matter theory. This line of reasoning, however, would be weird to say the least, since it pushes us to reject a well-defined and consistent physical theory \emph{based only} on metaphysical considerations. The other option would be to grant the theory physical dignity, in the sense that we should judge its validity based on its empirical adequacy. This second choice is potentially even worse, because it would drive us into something that smells like a case of empirical incoherence:\footnote{The definition of empirical incoherence given here is taken from \citet[][section 1]{314}, who follow in turn \citet[][section 4.5.2]{204}.} the truth of this theory would undermine our empirical justification for believing it to be true. This is because, if induced matter theory fulfills the GR-desideratum, there is no way to account for whatever measurement in a $5$-diffeomorphic invariant way.\\
Such a situation is all the more strange for the following reason. Since induced matter theory rests on the Campbell-Magaard theorem, GR is embedded in such a theory already at the mathematical level, which means that the former is a formal generalization of the latter and hence, if we forget for a moment the GR-desideratum, all the empirical tests that corroborate GR, would corroborate induced matter theory. Moreover, the theory would give novel testable predictions, mainly related to deviations in the geodesic motion of $4$-dimensional objects due to the presence of the uncompactified fifth dimension (see, e.g., \citealp{54}).\\
In order to find a solution to the problem of matter compatible with the GR-desideratum, some coherent account could in principle be put forward that saves the reality of $4$-dimensional material entities (including measuring devices), perhaps by appealing to some distinction between ``partial'' and ``complete'' observables \`a la Rovelli (\citealp{67}), but that would look more like a patch rather than a real solution. Of course, we can prevent the problem from happening by putting in the theory $5$-dimensional matter, thus retrieving a $5$-dimensional analog of Einstein's field equations (\ref{efe1})/(\ref{efe2}). However this would not count as a genuine solution because (i) it would just amount to dismissing induced matter theory in favor of another one, and (ii) it would not straightforwardly wash away any ``perspectival'' character from $4$-dimensional matter: such a ``full'' $5$-dimensional generalization of GR would for example fire up a metaphysical debate between $5$- and $4$-dimensionalists that would be very similar to that between $4$- and $3$-dimensionalists in standard relativistic physics (for an example of the literature, see \citealp{396,397}).\\
Be it as it may, the moral to draw from the above analysis is that, in induced matter theory, the GR-desideratum has rather undesired consequences. Hence, if we are willing to solve the problem of matter without dismissing induced matter theory, we can do it in a straightforward way, i.e., by relaxing the GR-desideratum.\\
A possible strategy in this direction would be simply to add more structure to $5$-spacetime, for example a $5$-flow in spacetime. This would amount to selecting a privileged fifth-dimension-like vector field $\boldsymbol{\psi}$ with components $\psi_{A}\equiv (0,0,0,0,-1)$\footnote{Or, more precisely, such that the $ij$-part of $g^{AB}\psi_{A}$ is identically null.} by adding to the field equations (\ref{5fe}), the following ones:
\begin{subequations}\label{fol}
\begin{equation}
g^{AB}\psi_{A}\psi_{B}=-1,
\end{equation}
\begin{equation}
\psi_{A;B}=0.
\end{equation}
\end{subequations}
It is easy to see that $\boldsymbol{\psi}$ induces a privileged decomposition of the $5$-manifold $M_{5}$ in $4$-hypersurfaces that are normal to the vector field and, hence, spacetime-like: each $4$-slice of the foliation will have equation $\psi=const.$. For this reason, equations (\ref{fol}) restrict the possible topologies compatible with (\ref{5fe}) to $M_{5}=M_{4}\times\mathbb{R}$. In a sense, we are back to the case of the compactified Kaluza-Klein theory: we are using (\ref{fol}) as a dynamical constraint on the possible topologies. But there is more to that: (\ref{fol}) tells us that not only $M_{5}$ has to be foliable by spacetime-like hypersurfaces, but that there is, for each model, a \emph{distinguished} way to foliate it. In short, according to (\ref{fol}), $M_{5}$ is a ``pile'' of $4$-spacetimes.\\
Under this new reading, a $5$-model $\mathfrak{M}$ of the theory will not be just a couple $<M_{5},g_{AB}>$ but a \emph{triple} $<M_{5},g_{AB},\psi_{A}>$. Let us return to the two models whose line elements are (\ref{5min}) and (\ref{5frw}), which are related by the intrinsic transformation $f\in diff(M_{5})$ (dual to the extrinsic transformation (\ref{tru})): we see that they correspond to the new situation where the first model is, say, $<M_{5},g_{AB},\psi_{A}>$ and the second model is $<M_{5},f^{*}g_{AB},\psi_{A}>$. We immediately notice that, due to the introduction of the structure $\psi_{A}$, the two models represent \emph{distinct} physical situations unless $f^{*}\psi_{A}=\psi_{A}$.\\
Also in this case, we can be accused of cheating - not once but twice! Firstly because we are not considering the starting theory but a brand new one that admits a structure (namely, a privileged foliation). Secondly because we are hiding a metaphysically suspicious background structure behind (\ref{fol}).\\
As regards the first charge, the plea is: guilty. It is undeniable that the starting theory dealt just with a generic $5$-metric (\ref{indmet}) solution to the field equations (\ref{5fe}) and nothing more (e.g. no cilindricity or compactification conditions), but this is exactly at the root of the problem of matter: a $5$-dimensional theory just resting on (\ref{indmet}) and (\ref{5fe}) has simply not enough structure to ground claims about the existence of $4$-matter. Can a physical theory that places dynamical constraints so weak to risk to be empirically incoherent be considered really a theory in any useful physical sense? Under this reading, adding further structure to such a theory is more like \emph{completing} it, rather than merely changing it.\\
As regards the second charge, the plea is: not guilty. Although $\boldsymbol{\psi}$ technically is a background object as the Minkowski metric in (\ref{kg3}), still it cannot be charged of being metaphysically suspicious. While, in fact, it is correct to say that $\boldsymbol{\psi}$ affects the $4$-matter distributions in each model without being affected in return, still the ``influence'' we are talking about is not a physical interaction ($\boldsymbol{\psi}$ does not really push $4$-matter in any physical sense), but an ontological one: in each model $4$-matter supervenes on the $5$-dimensional structure comprising a $5$-metric $\mathbf{g}$ \emph{and} a vector field $\boldsymbol{\psi}$.\\
To sum up, it is true that $\boldsymbol{\psi}$ represents an additional and ``rigid'' structure besides the $5$-metric but, as in the Kaluza-Klein case, there is nothing particularly wrong in that as long as we do not mind relaxing the GR-desideratum. Indeed, the payoff for this relaxation is huge: the problem of matter just vanishes.

\section{Conclusion}
The starting point of this paper was considering the possibility that, although general covariance, diffeomorphism invariance, and background independence are not features uniquely ascribable to GR, there is something in the way GR implements such features that renders this theory better than its predecessors. We tentatively identified this ``something'' in the GR-desideratum, and we then discussed two cases in which forcing such a desideratum on theories that seek to generalize GR leads to conceptual difficulties. We finally suggested to overcome these difficulties just by relaxing the GR-desideratum, in particular by introducing further (dynamical) spatiotemporal structures besides the metric. From what has been discussed in section \ref{se1}, it appears clear that questioning the GR-desideratum is not questioning general covariance, diffeomorphism invariance, or background independence by themselves, but the necessity of implementing these features in the way GR does.\\
The analysis developed in this paper is far from being just an otiose conceptual exercise, since fully understanding what the GR-desideratum really is might shed light on two huge open problems in the philosophy and physics of spacetime theories. Namely, (i) the problem of time in Hamiltonian GR and (ii) the problem of constructing a theory that combines the main tenets of relativistic physics with the empirically proven non-locality of quantum theories.\\
The first problem (debated for example in \citealp{7,8}), roughly speaking, arises when we decompose the $4$-dimensional spacetime of GR by means of space-like $3$-surfaces in the so-called ADM formulation of Hamiltonian GR \citep{108}. The idea behind this (3+1) decomposition is that we can render the general relativistic dynamics simpler for calculational purposes by using the machinery of Hamiltonian dynamics. In such a setting, we specify a set of initial conditions on a starting $3$-surface and we evolve it using Hamilton's equations of motion. Of course, such a decomposition is merely formal: it is just a convenient way to recast the standard dynamics of GR. However, there is a conceptual problem lurking beneath the formalism.\\
To see this, let us consider some arbitrary model of GR; since the (3+1) decomposition is merely formal, we can imagine two different ways to foliate the model by means of the foliations $\mathcal{F}$ and $\mathcal{F}'$. Let us further assume that the foliations agree only on the initial surface $\Sigma_{0}$.
We are now in the strange situation already experienced in the previous section: even by specifying with arbitrary precision the initial data on $\Sigma_{0}$, the dynamics is unable to single out one evolution among $\mathcal{F}$ and $\mathcal{F}'$. Does this imply indeterminism? Of course no, because both evolutions represent one and the same physical situation. However, if we take the freedom to foliate the model as a gauge freedom, then we are forced to say that $\mathcal{F}$ and $\mathcal{F}'$ are physically indistinguishable, which means that the only physically meaningful observables definable in this context are those that do not change whatever foliation we choose. This excludes that whatever physically significant quantity changes in time: taking this picture seriously seems to imply that the universe is a frozen block. Note how this problem stems from forcing a strict requirement of gauge invariance on the theory: from this point of view, the problem of time in Hamiltonian GR is strikingly similar to the problem of matter in induced matter theory. Also in this case, then, a way out of the conundrum would be to loosen or modify the gauge requirement so that other observables besides the unchanging ones could be defined (a solution of the problem has been in fact proposed by \citealp{420}).\\
The second problem (discussed at length in \citealp{238,198}), can be roughly summarized as follows. Let us imagine a quantum system made up of space-like separated parts $A$ and $B$ in an entangled state, and let us consider an event $E(A)$ such as a certain outcome of a measurement performed on $A$. It has been experimentally shown \citep{248} that $E(A)$ is not determined solely by the events in its past light-cone. This is what Bell called non-locality \citep[][ch. 2]{195} and that might mean that $E(A)$ can be affected either by events in its future light-cone or by space-like separated events such as a measurement taking place on $B$. In any case, quantum non-locality seems to be at odds with the physical interpretation of the light-cone structure of relativistic spacetimes, so that some modification of such a structure might be called for.\\
If we consider that one of the most promising theoretical programs towards the construction of a quantum theory of gravitational phenomena (viz. canonical quantum gravity) seeks to quantize GR starting from its Hamiltonian formulation, we immediately realize that this program faces a conceptual tangle that is basically the combination of the above mentioned problems. This is a further reason for reflecting on the GR-desideratum and finding ways to relax it without giving up general covariance, diffeomorphism invariance, and background independence.

\pdfbookmark[1]{Acknowledgements}{acknowledgements}
\begin{center}
\textbf{Acknowledgements}:
\end{center}
I would like to thank J. Brian Pitts and an anonymous referee for the useful comments on an earlier draft of this paper. Research contributing to this paper was funded by the Swiss National Science Foundation through the research grant no. $105212\_149650$.

\pdfbookmark[1]{References}{references}
\bibliography{biblio}
\end{document}